# Raining lead around 250mya : a smoking gun for an Australian impact origin of the Permian Extinction.

(2170 words)

Jim Standard and C. Austen Angell
Department of Chemistry and Biochemistry, Arizona State University,
Tempe, AZ 85287

**Recent documentation of extreme atmosheric sulfur and methane coincident with the vast Permo-Triassic (P-T) extinction allows interpretation of a 40-year old report of metallic lead microspheres, with aerodynamic tails, in the graphite-loaded fluvial strata of early Triassic sandstones in the Sydney basin. While climate change and flood basalt volcanism could explain the atmospheric anomalies, only an extraterrestrial impact in a lead sulfide mineralized zone could explain the occurrence of native lead in this form. Using thermodynamic arguments, flow directional data for the sandstones, and Tasmanian mineralization data for the lead origin, we deduce an impact site in Bass Strait, where ring-like gravitational anomalies, and a provocative "interior basin" structure complete with melt-rocks at 2 km depth (as in the Cretaceous-Tertiary (K-T) boundary Chixculub crater) are found in gravity survey and oil drilling maps/reports. We predict the co-occurrence of lead tears and graphite elsewhere, possibly in Antarctic sandstones.**

Since the impact origin of the K-T boundary extinction [1] was established to the satisfaction of most [2], there has naturally arisen a great interest in the possibility that the even more catastrophic die-off at the P-T boundary has a similar origin. The possibility has been strengthened recently by the evidence from Rampino et al. [3,4] and others [5] that the boundary is extremely sharp on the geological time scale. Based on the study of Austrian alpine rocks at the P-T boundary, Rampino et al concluded that the period in which the extinction occurred was no more than 60,000 yr, perhaps less than 8,000 yr. Fungal studies by Steiner et al [4] in the Karoo Basin of South Africa suggest an interval less than 40,000 years (also a geological eye-blink). Coupled with suggestions [6,7] that non-terrestial isotope ratios of gases were encapsulated in buckey-balls (that can form intense heat spikes], these observations may be construed to imply that a large bolide struck the earth at that time. However, unlike the iridium-rich layer of clay that is found across the world at the T-C boundary, there has been no "smoking gun" found for the P-T boundary. Likewise, no convincing impact site, or even the usual shocked quartz impact markers [8] have been discovered [9]. On the other hand, in the Meishin beds of China that have been so well studied[10,11,12] and precisely dated at 251.4 ± 1 m.y. ago [13],.glass spherules [10], and particularly metallic Fe-(Si-Ni) grains [11] previously associated with impact events, have been reported

An observation that would constitute a "smoking gun" for a nearby impact event, if it could be proven free of any artifactual origin, is that of metallic lead microspheres (~0.5mm diam) in the early Triassic sandstones of the Sydney basin, Australia, reported 40 years ago by Bayliss and Standard [14], and ignored ever since. Unlike most sandstones, the Hawkesbury sandstones are quite rich in graphite [15]. Graphite grains are the most obvious non-quartz component of these quartz-rich sandstones. Although graphite is common in nature, it has also been associated with impacts sites [16] and interpreted in terms of the high temperatures required for its formation from less structured forms of carbon (either in the impactor itself or in the impact site). Graphite is found throughout the Hawkesbury Sandstone. Pieces as large as 1 cm in diameter have been recovered [15]. The low chemical potential of oxygen associated with its presence provides a plausible explanation for the persistence, through geologic time, of lead in its metallic state. Its persistence is much less surprising than that of the more electropositive FeSiNi alloy, an accepted impact marker, found in the South China P-T boundary layer in ref. 11 and more recently at Graphite Peak in Antarctica, along with other impact evidence [21]. Here we couple the metallic lead observation with other sets of observations from the present work [15], and previous observations on the early Triassic atmosphere [11, 17-19], to argue for a bolide impact in the Bass Strait region of Australia, where a tell-tale circular anomaly is identified from seafloor and gravity maps.

In Fig. 1 we show examples of the lead teardrops, and in Fig. 2 we provide quantitative support for the their occurrence, based on thermodynamic data [22]. Fig. 1 shows that, of the common metal oxides, PbO and CuO are those reduced to metal at the lowest temperature in the presence reducing gases. Fig. 1 also shows that the common lead mineral PbS decomposes to the elements if the temperature rises above 1000 ºC, as would happen in the heat of impact. In view of its high density and, in the present case, also greater abundance (see below), lead metal will be the first, and probably the only, metal to "rain out" of the cooling gases or collect in the proximal ejecta rim .

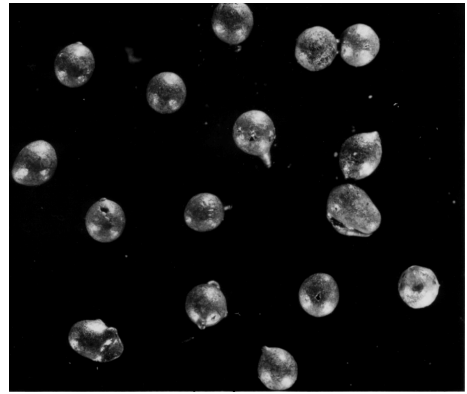

Fig. 1. A selection of the 165 lead microspheres, with tails, found in the Hawkesbury sandstones, and reported in refs. 14,15.

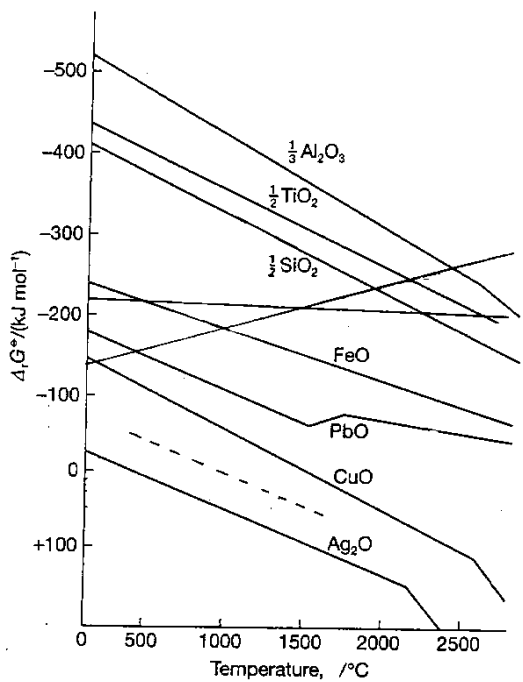

**Figure 2.** Ellingham diagram [22] for the free energy of formation $\Delta G°_f$ of metal oxides in relation to temperature. The negative slope reflects the entropy increase on decomposition due to the fact that oxygen is a gas. All divalent metal oxides have comparable slopes on this plot. There are two lines cutting across those of $\Delta G°_f$ for solid oxides. These are the free energies of combustion $\Delta G°$ of carbon to give carbon monoxide (positive slope) and of combustion of methane $CH_4$ to give gaseous $H_2O$ and $CO_2$. Their intersections with a given metal curve give the temperature, under standard conditions, above which graphite, or $CH_4$ respectively, will reduce the metal oxide to the metal. It is seen that, in an atmosphere rich in methane metallic lead will form from the oxide at much lower temperatures than for any other common metal except copper. Likewise, during a cooling process from high temperatures, lead and copper will be the last metals to recombine with oxygen (e.g. during descent from an apocalyptic atmosphere).

Fig. 2 also contains a plot for the sulfide of lead (dashed line). $\Delta G°_f$ is positive above ~1000°C, i.e. PbS will decompose thermally to lead and sulfur vapor at any higher temperature. Naturally occurring sulfide mineral deposits impacted by a bolide would be vaporized with decomposition to the elements and, in the presence of atmospheric oxygen, will form the abundant sulfur dioxide identified in the early Triassic atmosphere.

---

The lead microspheres were discovered by the sifting of disaggregated samples of sandstone which were mostly taken from surface outcrops [15], and we have no in situ cases to render their natural origin unambiguous. Also, no copper droplets have been found. Thus we must give some consideration to possible artifactual origins (particularly as the lead is of terrestial isotope abundance (age $10^9$ yr) and is depleted in silver [14]). One that comes immediately to mind is debris from the shot gun blasts of bird and animal hunters. Lead shot fragments could conceivably reach melting point by freak frictional encounters with nearby rocks and hence, during air passage, to develop the tear drop character found in the recovered spheres. We have made efforts to produce such debris from shotgun blasts, but have had no success. The uniformity of size (0.5 mm diam), as well as the teardrop shape seen in Fig. 1 [15,23], favors a vapor condensation origin. While rare in Nature, metallic lead in spheroidal form has been found elsewhere (in an ocean drill core, located close to the K-T boundary) by Jon Haglund of USGS (private communication). In the following we assume that the lead microspheres are naturally occurring, and follow this to an interesting conclusion that is now receiving study in the field.

The Hawkesbury Sandstone lies atop other early Triassic sandstones, that are generally described as the Narabeen

Formation. At the New South Wales coast (Coalcliff), the Narabeen Formation lies unconformably on Permian coal. The dividing layer, which is a 8-10cm black claystone breccia, has been identified by Retallack [17,18] as the Permian-Triassic boundary layer, on the basis of the great loss of biological diversity, and high level of fungal spores, in the layers above. The description "black claystone breccia" has also been applied recently to the better known P-T boundary layer in South Africa's Karoo Basin [19]. Accepting this identification of the location of the 251.4 m.y. P-T event, the Hawkesbury sandstone must be of more recent origin, by an interval of some 5 m.y. according to Retallack [17] though the dating seems to be rather uncertain. The presence or otherwise of lead microtears in the Narabeen Formation has not been properly investigated (although two occurrences of the lead microtears in the ref. 15 study were in Narabeen rocks).

The direction of flow during deposition is indicated by the thickness contours of the formation shown in Fig.3 and is supported by detailed cross-bedding analysis [15] the material for the graphitized sandstones came from the southwest, superimposing on an existing filled basin (the Permian coal lies hundreds of feet below Sydney under the Narabeen formation. The Hawkwesbury Sandstone thickens to the northeast to a present day maximum of nearly 300 m near the mouth of the present Hawkesbury river. Retallack [17] describes early Triassic deposits that came from the northwest but, on the basis of heavy mineral content, these are quite distinct from the Hawkesbury formation of the ref. 15 study.

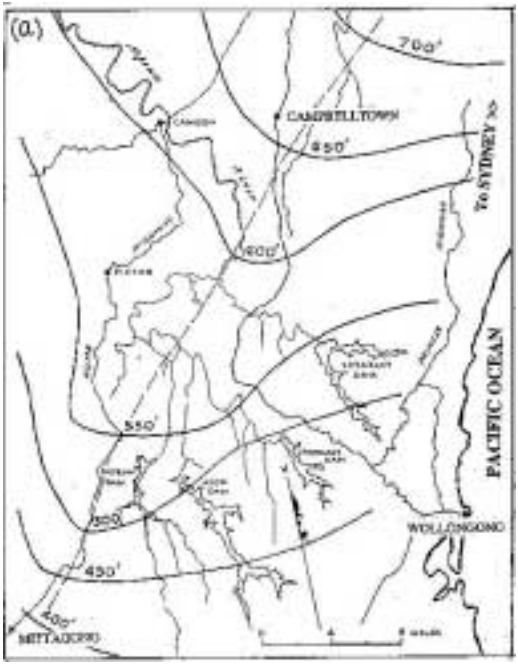

**Figure 3.** Contour map of the total thickness of the lead and graphite-carrying Hawkesbury sandstones of the southern part of the Sydney basin, showing thickness increasing in the direction North-north-east. These beds are believed to have been laid down, over earlier Triassic sediments, within 6 m.y. of the P-T extinction. (From ref. 15)

Supposing, then, that the lead spherule-preserving graphitized sandstone is indeed closely linked in origin to a bolide impact, we conclude that the impact site must lie far to the south of the Sydney basin. The largest graphite sample (~1 cm diam) was found near the southern extremity of the formation where there would have been least opportunity for flow abrasion. The Hawkesbury sandstone terminates abruptly at an escarpment caused by erosion from the more recent Shoalhaven river system that was created by the Tertiary uplift that formed the Australian Alps.

At the time of Permian extinction, the whole of south-eastern Australia was a flat granite plain. Assuming that impact-derived components of the Hawkesbury Sandstone source material fell on this plain (e.g., the vicinity of the present New South Wales/Victoria boundary) then the considerations of proximal ejecta distribution given by Melosh [24,25] would suggest an impact site some 4-5 crater diameters to the south [8,24,25]. To locate this site another indicator line is needed.

Since we have argued for a sulfide mineral deposit impact site origin for both the lead and the sulfur dioxide that suddenly became so abundant at the P-T boundary [11], it is natural to seek the trajectory of the nearest major sulfide mineralization region. This is located in western Tasmania where extensive lead/zinc ore deposits (with proportionately very little copper) occur in a Cambrian volcanic mineralized belt. Examining the number density of lead/zinc mines in Tasmania (see Supplementary Material) we obtain a rough ribbon running North-easterly across the Tasmanian mainland and projecting out into Bass Strait as indicated in **Fig. 4.** Bass Strait, 250 m.y. ago, may have been a shallow sea area rich in the methane hydrates that have been proposed by others [18,26] as a source of impact-generated [19] or other-cause anoxia [27] involved in the P-T extinction. Additionally, or alternatively, the area contained Permian coal seams that occur in Tasmania on either side of the mineralized region. Both would be sources of the light carbon needed to interpret carbon isotope ratio findings at the P-T boundary - which Siberian volcanic $CO_2$ cannot [19]. At the time of impact, Permian coal would have been in the form of peat and could have provided the source of carbon for the hydrocarbon gases (primarily methane in the intense heat of impact) in the atmosphere at that time [19-20], and of the graphite in the sandstones.

Support for our sulfide-mineralized impact site hypothesis comes from the report of Kaiho et al [10] on the P-T boundary in southern China, where they show that the enrichment ratio of the element copper in the bed no. 24-3 (small grain component), greatly exceeds that of nickel. This anomaly was passed over without comment in ref. 10. No data for zinc were reported. We would expect a zinc anomaly like that for copper but larger. Kaiho et al emphasized the "gigantic release of sulphur" based on the sulfur isotope excursion observed in the uppermost Permian limestone bed. This led them to conclude that the oceanic sulfur content briefly doubled at the P-T boundary A sulfide-mineralized impact site would obviously generate huge quantities of $SO_2$ gas that would quickly become sulfuric acid. Indeed Kaiho et al [11] have suggested that the P-T extinction was caused the generation of some $4 \times 10^{19}$ moles ($10^5$ billion tons) of $H_2SO_4$ in the atmosphere so that the ocean pH briefly became like vinegar.

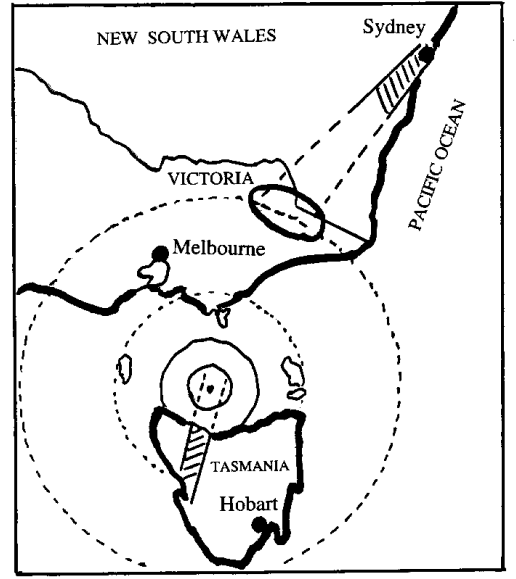

**Figure 4.** Map of South-Eastern Australia and Tasmania, showing lead/zinc sulfide mineralized belt of Tasmania projected into the middle of Bass Strait, and the proposed crater site (based on gravitational anomaly in Fig 5) The 90% ejecta limits for 100 km and 200 km diameter craters, according to Melosh [24,25] and French [8], are shown by dashed circles. The overlap with the suggested source area for the lead and graphite-containing Hawkesbury sandstones of the Sydney basin (upper right), is indicated by an elipse

The ecological devastation downwind from heavy metal sulfide smelting sites is well known from modern times. That an impact instantaneously smelt-vaporizing a region equivalent to the entire mineralized region of Tasmania, should cause such widespread and immediate devastation both on land [19] and in ocean [10], is not difficult to understand. In view of the toxic character of the impact products, the impact need not have been as huge as is often supposed in order to account for the extent of the extinction.

The above indicators from the Sydney basin deposits and the Tasmanian mineralized belt are combined in Figure 4 to suggest an impact site in Bass Strait between Victoria and Tasmania. Indeed, we find remarkable aspects to the geological structure of this shallow marine shelf, which we now discuss.

Firstly we note the depth contours in Bass Strait that are reproduced, in Fig. 5a, from ref. 28. There is a roughly circular structure of depth never greater than 80m with a central feature (arrow in Fig. 4a) reminiscent of the central uplift peak caused by bounce-back of the most highly compressed zone at the center of large magnitude events [8]. Secondly we observe that the central feature in Bass Strait correlates with an impressive gravitational anomaly of circular form seen at the center of Fig. 5b [29]. A similar circular gravitational anomaly was one of the clues leading to the discovery of the Chixculub crater in Yucatan, and gravity contours are also the source of the recognition [2]. that the Chixculub impact structure is much larger than originally supposed

These observations suggest the study of reports from oil explorations of the Bass Strait area (conducted in the period 1960-75). We find the following description [30] (see Supplementary Material) of the Bass Basin, which is roughly symmetrical around the gravitational anomaly of Fig 4b: "The Bass basin is technically an interior basin, *being surrounded on all four sides by older continental rocks"* (italics ours). The bore holes reportedly penetrated 7000 ft of "sediments" before encountering, in two cases near the suggested impact center, "volcanic rock". The "volcanic rock", (descriptive of a quick-frozen melt) occurs at roughly the same depth, ~2 km, as the impact melt plate in the Chixculub crater (see Fig. 4 in ref. 2). Fortunately these drill cores are available for more detailed study. Until such studies have been carried out our interpretations must remain in the category of informed speculation.

In the light of these observations, there are several locations along the north coast of Tasmania that would seem to warrant closer examination. For instance, near Ulverstone, early geological maps identify a two mile stretch of coastline as Megabreccia, and also list a location called Washbrook Chaos. One of us (JCS), currently in the field in Tasmania, has identified shatter cones [8] amongst the Megabreccia rocks, and specimens are being accumulated for microscopic examination. We expect these micrographs to contain the compelling planar deformation features (PDF s) [8] needed to confirm a nearby impact site. Shocked grains may be identified among the photomicrographs from ref. 15, but are not abundant.

## Acknowledgements.

We wish to acknowledge helpful, critical, comments on this work from George Wolf and Peter Buseck, and the untiring support of Cliff Kwan-Gett.

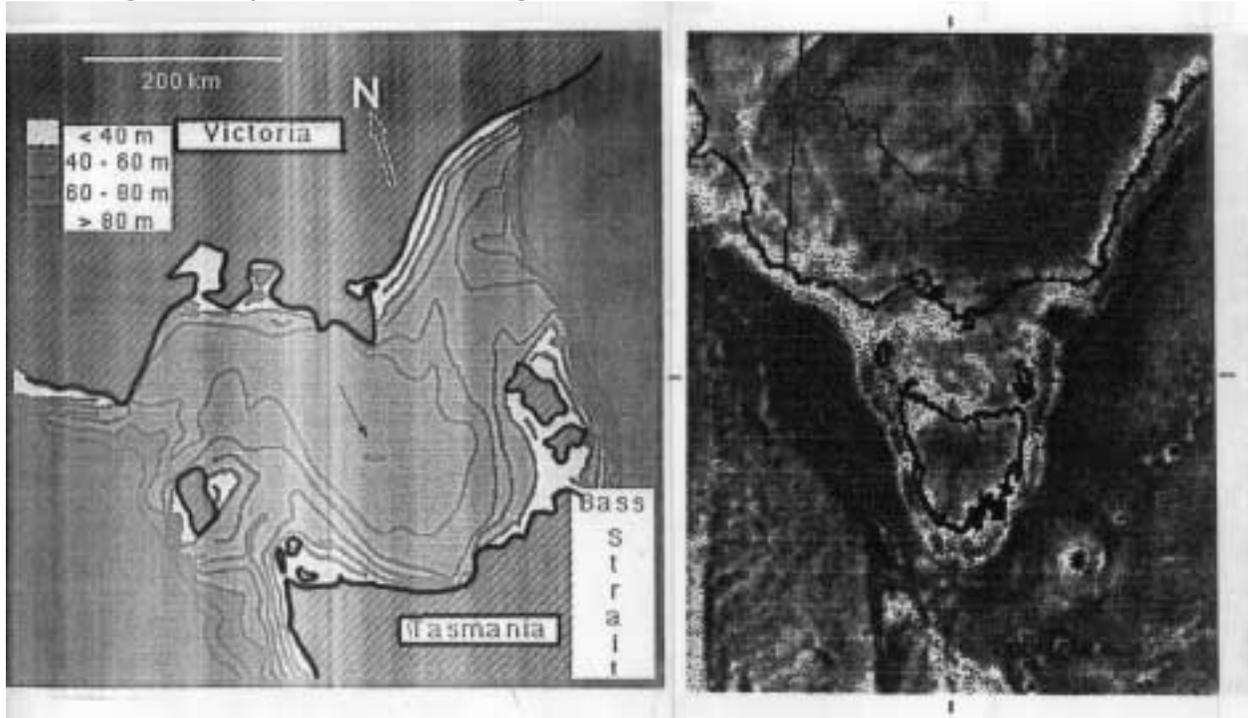

**Figure 5.** (a) Depth contour map of Bass Strait, between Victoria and Tasmania, showing possible uplift peak affecting sea floor sediments in the generally flat sea bed. This is found at the same point as the ring-like gravitational anomaly in part (b).

(b) Gravitational anomaly (bright spot in center of dark ring, at mid-figure - use cross-hairs) corresponding to the depth contour peak of part (a). The right arc of a second dark ring may also be discerned. The center of this anomaly was used as the impact point in the construction of Fig. 3b.